\documentclass[showpacs,aps,prb,twocolumn,floatfix,superscriptaddress]{revtex4}
\usepackage{CJK}   
\usepackage{graphicx}
\usepackage{epsfig}
\usepackage{wasysym}
\usepackage{color}
\usepackage{hyperref}
\newcommand{\sumnear}{\mathop{\sum}_{\langle i j \rangle}}

\begin{document}
\begin{CJK*}{GBK}{}
\title{Static impurities in a supersolid of interacting hard-core bosons on a
    triangular lattice}
\author{Xue-Feng Zhang}
\email{zxf@itp.ac.cn}
\affiliation{Physics Dept.~and Res.~Center
OPTIMAS, Univ.~of Kaiserslautern, 67663 Kaiserslautern, Germany}
\affiliation{Institute of Theoretical
Physics, Chinese Academy of Sciences, P.O. Box 2735, Beijing
100190, China}
\author{Yu-Chuan Wen}
\affiliation{Center of Theoretical Physics,
Dept.~of Physics,
Capital Normal Univ., Beijing 10048, China}
\author{Sebastian Eggert}
\affiliation{Physics Dept.~and Res.~Center
OPTIMAS, Univ.~of Kaiserslautern, 67663 Kaiserslautern, Germany}


\begin{abstract}
We study the effect of impurities in a supersolid phase  in
comparison to the behavior in the solid and superfluid phases.
A supersolid phase has been established for 
interacting hardcore bosons on a triangular lattice which may be realizable by 
ultracold atomic gases.
Static vacancies are  considered in this model which
always lower the magnitude of the order parameter in the 
solid or superfluid phases.
In the supersolid phase, however, the impurities
directly affect both order parameters simultaneously and thereby reveal an interesting
interplay between them.
In particular the solid order may be enhanced at the cost of a
strong reduction of the superfluidity, which shows that the
two order parameters cannot be in a simple superposition.
We also observe an unusual impurity pinning effect in the solid ordered phase, which
results in two
distinct states separated by a first-order transition.
\end{abstract}



\maketitle
\end{CJK*}

A bosonic supersolid phase is characterized by the coexistence of
two seemingly contradictory
order parameters, a solid crystalline order and a superfluid density.
This reflects the spontaneous breaking of two independent
symmetries, namely translation and a U(1) gauge rotation,
which are also known as diagonal and
off-diagonal order, respectively.
The simultaneous breaking of two independent symmetries
in the supersolid phase is counter-intuitive and unusual,
because normally a spontaneously broken order
locks the system into a single phase.
Only when the remaining fluctuations are large enough, two independent order parameters
may exist in one phase, e.g.~due to frustration.
Having been predicted forty years ago,\cite{supersolid1}
supersolids recently received
renewed interest after a possible observation in $^4$He.\cite{supersolid_experiment1}
The presence of mobile $^3$He impurities appears to be important in those systems, which 
are predicted to raise $T_c$ but reduce the superfluid density.\cite{He3}
While the experiments are still controversially discussed,\cite{supersoliddoubt}
there is now very strong numerical evidence that a supersolid phase
is realized for interacting hardcore bosons on a triangular
lattice.\cite{tri1,tri2,tri4,tri6,tri7,tri8,tri9}
{Such a model can potentially be realized by ultracold atoms in optical traps.\cite{Pollet}
Tunable superexchange models have already been
experimentally created in this rapidly emerging field\cite{SXinOL} and
interacting hard core bosons\cite{perturb} have also been discussed.
Most recently also triangular lattices have been possible\cite{sengstock} so 
it appears likely that supersolidity will soon be a central topic for 
hard core boson models in the ultra-cold gases community.}

While the coexistence of the corresponding two order parameters 
is well established numerically in the hard core 
boson systems,\cite{tri2} the
microscopic interplay between them is still unclear.
We now 
study the
impurity effects on both order parameters simultaneously in the supersolid phase,
in order to clarify if the two order parameters are in a simple superposition
or how they may interact locally. 
The use of substitutional impurities in strongly correlated systems
has become a standard tool for understanding the underlying
quantum phases.\cite{imp1,imp2,imp3,imp4}
In particular, it is possible to study local
expectations values around defects\cite{imp5,imp6} 
for an analysis of the elementary excitations and
direct comparison with theoretical models.
In the supersolid phase we are now able to 
consider the effect of static impurities on two coexisting order parameters
simultaneously.

The model we will consider in this paper is the spin-1/2
model on a triangular lattice
\begin{eqnarray}
H=-t\sumnear(\hat{S}_{i}^{+}\hat{S}_{j}^{-}+ h.c.)
+V\sumnear
\hat{S}_{i}^{z}\hat{S}_{j}^{z} -B\mathop{\sum}_i
\hat{S}_i^z,\label{XXZ}
\end{eqnarray}
with antiferromagnetic exchange $V$ in the z-direction corresponding
to nearest neighbor repulsion and ferromagnetic exchange $t$ in the
x-y-directions, corresponding to the kinetic energy and $B=\mu-3V$
in terms of the chemical potential of the equivalent hard-core boson
problem.  The simplest
impurities are given by lattice vacancies in model (\ref{XXZ}).\cite{remark} 

The two order parameters in the supersolid are given
by the structure factor $S(q\!\!=\!\![4/3\pi,0])=\langle
|\mathop{\sum}_{r} \hat S^z_r e^{i q \cdot r} |^2\rangle/N^2$ for
the solid order, and by the superfluid density $\rho_s$, which is
typically measured using the winding number $W$ in quantum Monte
Carlo (QMC) simulations $\rho_s=\langle W^2\rangle/4\beta t$.\cite{tri1,tri2,tri4,tri6,tri7,tri8,tri9,w} We use a modified
perturbation theory and the directed-loop stochastic series
expansion QMC algorithm\cite{sse1} { with finite size scaling up to $N=324$
 sites at a temperature of $T=0.02V$. }
In order to avoid trapping in one of the degenerate states
it is also essential to implement parallel tempering
in the parameter space.\cite{para1,para2}

\begin{figure}
\includegraphics[width=0.5\textwidth]{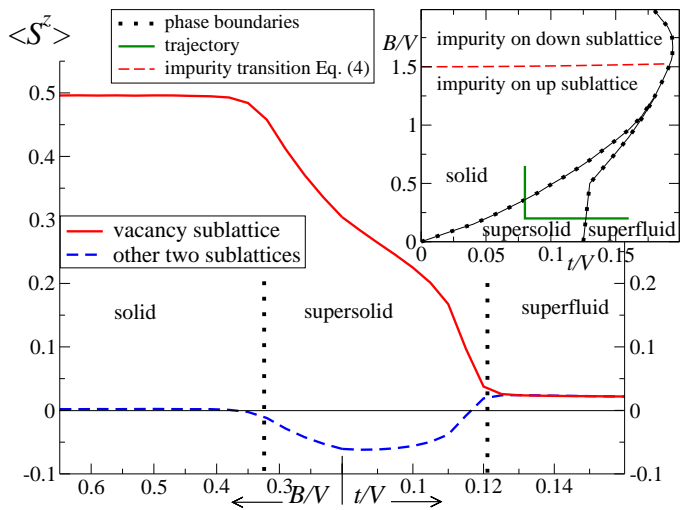}%
\caption{(color online). The average magnetic moment of the
sublattice  which  contains the vacancy (solid) and the other two
sublattices (dashed). Inset: The phase diagram with the impurity
phases and the plot trajectory we use in Figs.~\ref{rho},
\ref{sqrhos}, and \ref{multi-imp} along $t/V=0.08$ and then along
$B/V=0.2$.
 \label{rho}}
\end{figure}

For reference we first examine a single vacancy in the solid phase, 
which already shows interesting effects.
As shown in the inset of Fig.~\ref{rho}
this phase occurs for small xy-coupling $t$.
The solid order is characterized by a $2/3$ filled state
for positive fields
with exactly two spins on each triangle pointing up.
 For negative fields there is an equivalent $1/3$ filled ordered phase
due to the spin flip symmetry around $B=0$.
The vacancy in the XXZ model (\ref{XXZ}) does not break this
symmetry,\cite{remark} so it is sufficient to
consider only positive fields 
$B>0$ in the phase diagram in Fig.~\ref{rho}.
The choice of the spin-down sublattice (pointing against the field)
gives a three-fold degeneracy,
 which is however lifted by the vacancy.
In particular, for $0<B<1.5V$ and $t=0$ the vacancy site must belong to
one of the spin-up sublattices,
as can be seen by simple energetic considerations.
Therefore the order in the entire system is pinned by a single defect
and only a two-fold degeneracy remains.
The average occupation $\langle S^z\rangle$ on the different sublattices in
Fig.~\ref{rho} shows that this pinning also continues throughout
the supersolid phase.
However, the spin density of the other two sublattices surprisingly point {\it against} the 
field in the supersolid phase.

For larger fields $B>1.5V$
there is a transition to a different state,
where the order is now pinned on the opposite sublattice  
with no remaining degeneracy.
 Therefore, a single impurity can in fact induce a  transition between two 
distinct ordered states of the entire system.
The transition line 
also depends on the xy-coupling $t$ as can be seen by perturbation
theory in the ''hopping'' terms $H_{ij}=-t(\hat{S}_{i}^{+}\hat{S}_{j}^{-}+h.c.)$.
Unfortunately, the usual perturbative correction to the wavefunction
$|\psi\rangle \approx |0\rangle + \sumnear |ij\rangle\langle ij|H_{ij}|0\rangle/(E_0- E_{ij})$
diverges with the number of hopping terms, i.e.~the system size $N$.
Here $|0\rangle$ is the ordered
state and $|ij\rangle$ has opposite spins exchanged on the bond $i,j$
relative to $|0\rangle$.  Of course the number of lattice sites $N$
must be irrelevant in the ordered phase,
so the trick is to modify the perturbation correction to include only those hopping terms
which actually affect a local expectation value.\cite{perturb}
For example, to calculate
the energy correction at one bond $\delta E_{ij}$
only the corresponding hopping term is considered
\begin{equation}
|\psi_{ij}\rangle \approx |0\rangle + |ij\rangle\langle ij|H_{ij}|0\rangle/(E_0- E_{ij})
\label{per1}
\end{equation}
and we simply get
\begin{equation}
\delta E_{ij} = |\langle ij| H_{ij}|0\rangle|^2/(E_0- E_{ij}) = t^2/(E_0- E_{ij}).
\end{equation}
For the case of a vacancy, the excitation energy 
$E_{ij}$ depends on the location of the bond and it also depends on which
sublattice is pointing down in the ordered state $|0\rangle$.
After summing over all contributions, we find that the energy difference
between the two possible pinned ordered states is given by
$\Delta E = B -\frac{3V}2-\frac{7t^2}{10V} +{\cal O}(t^3)$.
Therefore, the impurity driven transition line runs along
\begin{equation}
B\approx  3V/2 + 7t^2/10V \label{trans}
\end{equation}
as shown in the inset of
Fig.~\ref{rho}, which also agrees with our numerical QMC results.

\begin{figure}
\includegraphics[width=0.5\textwidth]{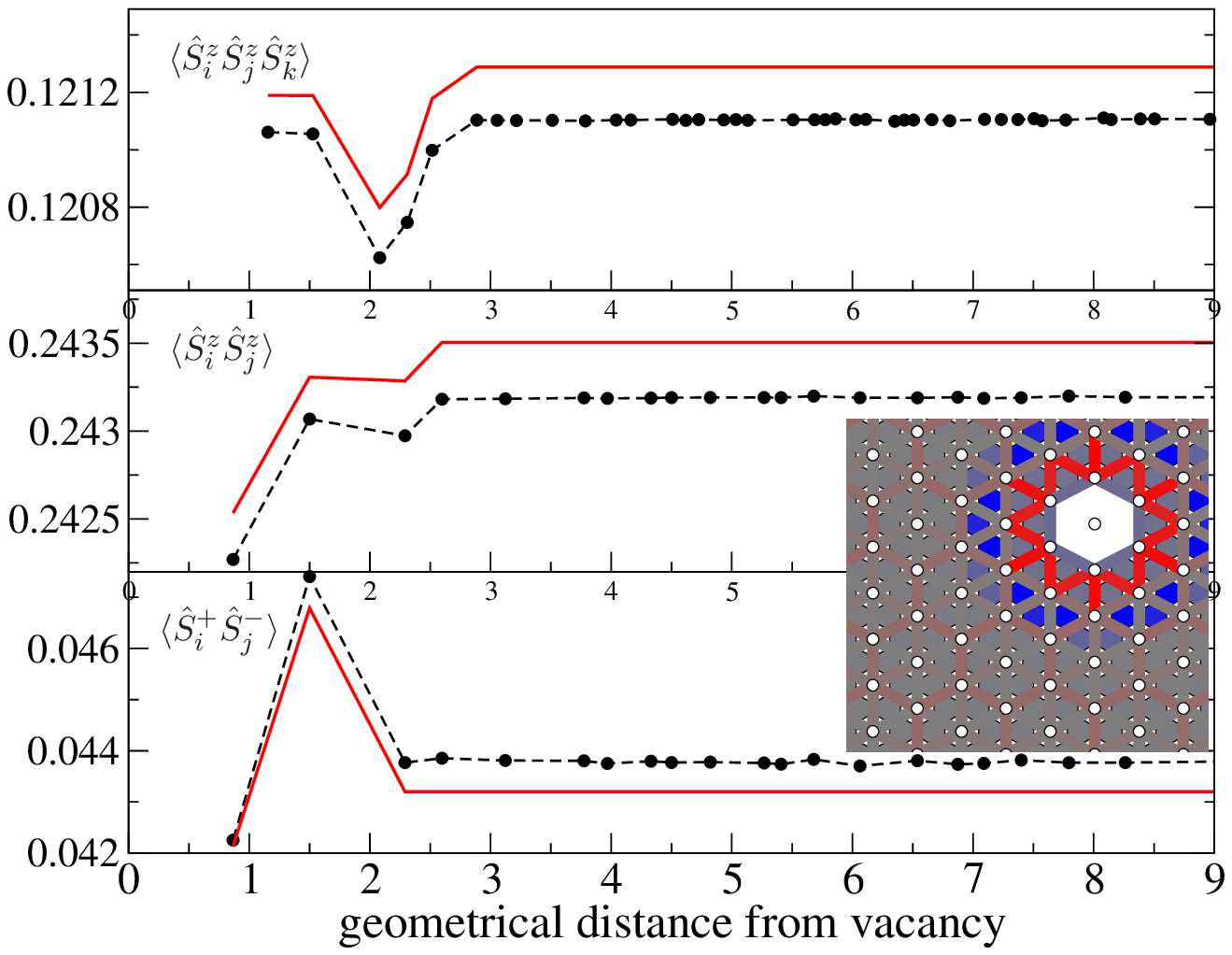}%
\caption{
(color online).
(a) $\langle \hat S^z_i \hat S^z_j \hat S^z_k\rangle$ with
$i,j,k$ on one triangle, (b) $\langle \hat S^z_i \hat S^z_j \rangle$
and (c) $\langle \hat S^+_i \hat S^-_j \rangle$, where $i,j$
are neighboring sites on sublattices not occupied by the impurity
as a function of distance in the solid phase ($t/V=0.08$, $B/V=0.65$, $N=144$).
The modified perturbation theory (solid) agrees very well  with
the QMC results (dashed).
Inset: Distribution of
$\langle \hat S^+_i \hat S^-_j \rangle$ and $\langle \hat S^z_i \hat S^z_j \hat S^z_k\rangle$ on the bonds and triangles, respectively.  Red signals an increase and blue a decrease relative to the gray bulk values.}
\label{sd}
\end{figure}

Using the modified perturbation theory with a restricted sum in Eq.~(\ref{per1})
it is also possible to analytically calculate local expectation values, e.g.~when calculating $\langle \hat S^z_i\rangle$ all hopping terms connecting to the site $i$ are included.
The results give a good indication about the
local order around the vacancies.
In particular, $\langle \hat S^z_i \hat S^z_j \rangle$ and
$\langle \hat S^z_i \hat S^z_j \hat S^z_k\rangle$ with $i,j,k$ on neighboring sites
are good indicators of the local solid order, which are reduced around the
vacancy.
On the other hand the
quantum fluctuations $\langle \hat S^+_i \hat S^-_j \rangle$ on bonds are enhanced as
shown in Fig.~\ref{sd}. Interestingly, the order reduction is {\it not} correlated in space
with the fluctuation enhancement and the effect is also not always strongest
directly at the vacancy.
The inset in
Fig.~\ref{sd} shows the distribution of the local expectation values around the impurity
on the lattice directly.  The expectation values
$\langle \hat S^+_i \hat S^-_j \rangle$ between two non-impurity sublattices is
different from those bonds involving the impurity sublattice even very far from
the impurity.  This is a secondary effect from the pinned order and should
not be mistaken for an independent  bond order.

\begin{figure}
\includegraphics[width=0.5\textwidth]{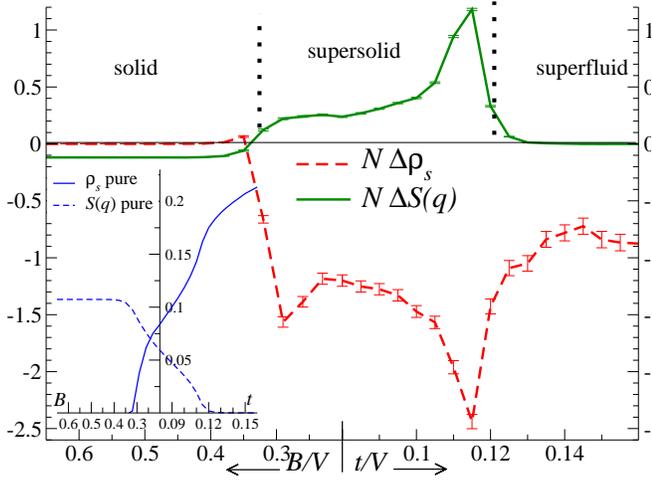}%
\caption{(color online). Scaled impurity corrections to
the superfluid density $N\Delta \rho_s$ and the
structure factor $N \Delta S(q)$.
Inset: Order parameters in a
pure system.  }
 \label{sqrhos}
\end{figure}

It is not surprising that
an impurity generally reduces the order parameter locally.
The main question for the supersolid phase
is now if the vacancy reduces {\it both} order parameters
as may be expected for a simple superposition of the
two effects or if  an interesting interplay can be observed.
The answer to this question is summarized in Fig.~\ref{sqrhos}, where we plotted
the impurity contributions of the two relevant order parameters in the system
as we cross the phase boundaries along the trajectory in the inset of Fig.~\ref{rho}.
The dominant parameter in the phases
with one single order is always reduced, while the other parameter
remains unchanged close to zero.  However, in the supersolid phase
only the superfluid density is strongly reduced, while
the solid order is in fact {\it enhanced}.
It is far from obvious why the vacancy
should enhance the solid order in this case, contrary to what we observed 
in the solid phase.
The only explanation of the observed behavior is that
the vacancy reduces the superfluid density locally so
strongly that the solid order is revived, which is evidence for
a microscopic competition between the two order parameters.
This result clearly shows that the two order parameters
are not in an independent superposition.
It is noteworthy that the competition and the total change of the order parameters
is strongest close to the second order phase transitions to the  
superfluid phase.

\begin{figure}
\includegraphics[width=0.5\textwidth]{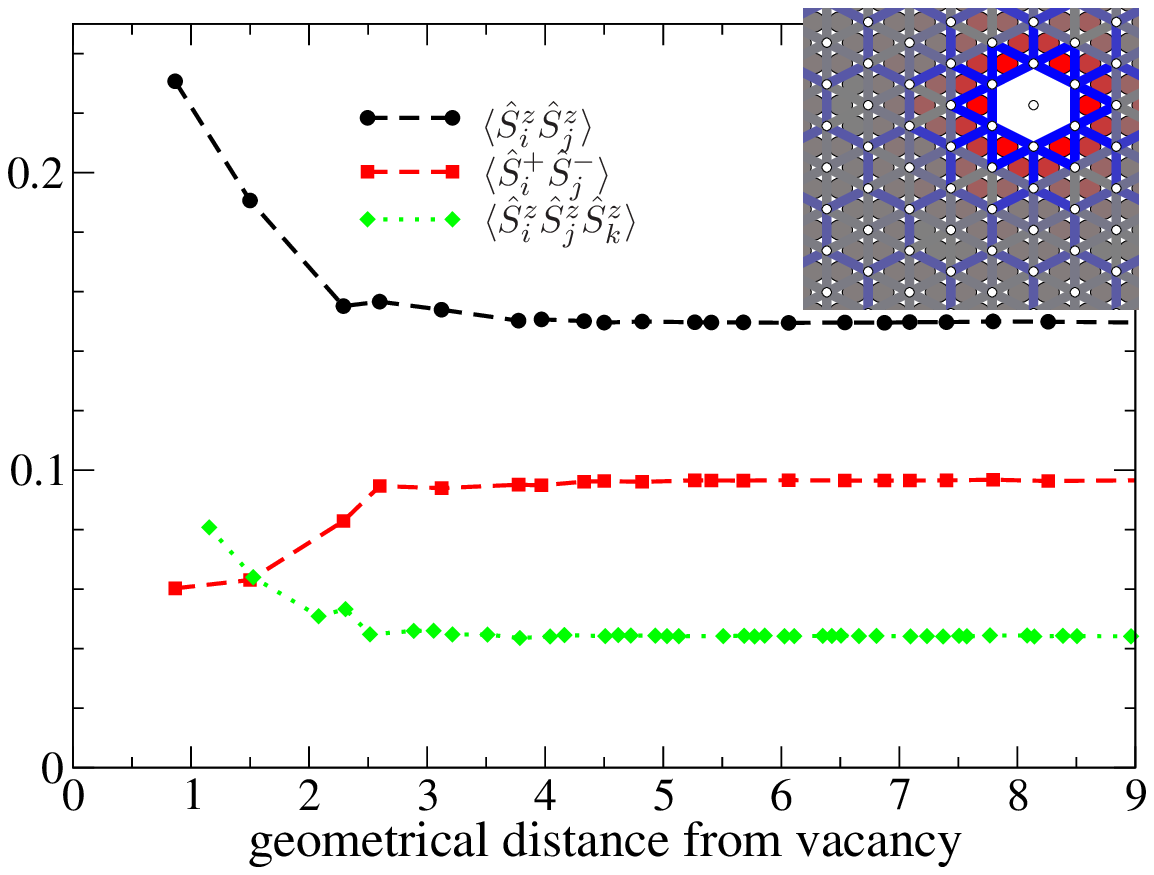}%
\caption{(color online). Local
expectation values as a function of geometrical distance analogous to Fig.~\ref{sd}
in the supersolid phase ($t/V=0.08$, $B/V=0.2$, $N=144$). Closest to the impurity the
values almost recover the bulk values in the solid phase in Fig.~\ref{sd}.
Inset: Analogous to inset~\ref{sd}, but showing the opposite behavior on a relative color scale.}
\label{ss}
\end{figure}

The local expectation values 
in Fig.~\ref{ss}  also demonstrate
the competition of order parameters locally.
The local solid order $\langle \hat S^z_i \hat S^z_j \rangle$
and $\langle \hat S^z_i \hat S^z_j \hat S^z_k\rangle$
close to the impurity is now enhanced
while the kinetic energy
$\langle \hat S^+_i \hat S^-_j \rangle$ is strongly reduced.
This is in strong contrast to the observations in the solid phase in Fig.~\ref{sd}
and the relative changes are also much more dramatic and correlated in space, which
again demonstrates the direct interplay between both order parameters.

\begin{figure}
\includegraphics[width=0.5\textwidth]{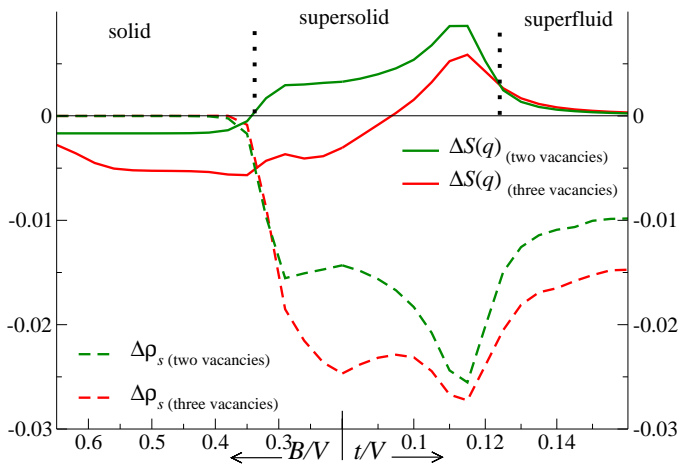}%
\caption{The impurity contributions to the structure factor and the superfluid
density for two and three impurities on different sublattices in a system of 144 sites.}
\label{multi-imp}
\end{figure}

We finally turn to the interesting case of several impurities in the
system.   A second vacancy on the same sublattice is rather
trivial, corresponding to the same pinned order, i.e. constructive interference
of the induced magnetization density.  Remarkably
also a second impurity on an opposite sublattice
has the same effect since the 
second impurity simply lifts the remaining two-fold degeneracy
exactly in such a way that both impurities are located on spin-up
sublattices.  The order is now completely pinned, but all observed effects
are approximately additive, i.e. the impurity contributions
to the order parameters simply double in the entire parameter space in Fig.~\ref{multi-imp}.  Also the observed phase transition
between pinned states in Fig.~\ref{rho} remains unchanged.  
This situation is in sharp contrast to two impurities on different sublattices
in an unfrustrated square lattice,\cite{imp6} which must show destructive interference
of the alternating magnetization.

The most interesting case is given by three impurities on
the three different sublattices.  The threefold degeneracy of the lattice
is then approximately restored again without pinning and the impurities must interfere
destructively.  This results in a surprisingly strong reduction of the
solid order $S(q)$ which now also carries over into the supersolid phase
as shown in Fig.~\ref{multi-imp}, while also
the reduction of $\rho_s$ remains  strong.
Obviously the effects are not simply additive in this case and indicate some
interesting impurity-impurity interactions.  However, in generic systems impurities 
break the 
symmetry between the three sublattices, so that the observed pinning 
and order parameter competition described above is the more general scenario.

In summary, we have used a modified perturbation theory
and QMC simulations to analyze impurity effects in
a supersolid in comparison to other phases with single order
as realized by the model in Eq.~(\ref{XXZ}).

In the solid phase a non-trivial pinning of the entire
order by a single defect has been observed.  Therefore, impurities 
create a 
first order transition line between two different pinned states
given by Eq.~(\ref{trans}), which is not seen in the pure system.  

In the supersolid phase the solid order is surprisingly enhanced by an impurity,
which coincides with a strong reduction
of the superfluid order.  This is evidence for an interesting
 microscopic competition
between the two order parameters, which certainly cannot be in
a simple superposition.  

For two impurities a simple addition of the observed effects can
be seen, while for three impurities on different sublattices
a strong destructive interference changes the physics completely. 
For a more complete understanding of the impurity-impurity interactions
on three different sublattices 
more research is needed.

In all impurity configurations
a very strong reduction of the superfluid
density $\rho_s$ occurs close to the second order supersolid-superfluid transition. 
It is therefore likely that the superfluid order
can be destroyed with a critical density of impurities, while the solid order
may be enhanced.
The extreme limit of this effect corresponds to the removal of one sublattice,
i.e.~the honeycomb lattice, which indeed
results in an extended solid phase.\cite{wessel}

The authors would like to thank H.G.~Evertz and J.Y.~Mao for helpful suggestions
and Yue Yu for great support and useful discussions.
This work was supported by the National Natural Science Foundation
of China under Grants No.10904096 and No.10604024, the Natural
Science Foundation of Beijing under Grant No.1092009, the DAAD,
and the DFG via the Research Center Transregio 49.

\bibliographystyle{apsrev}

\begin{thebibliography}{10}
\bibitem{remark}
Vacancies in the hard-core boson model are
equivalent only up to an additional symmetry-breaking
local field, which is a non-essential complication that is not considered here.
\end{thebibliography}

\end{document}